\documentclass[twocolumn]{aastex631}
\usepackage[T1]{fontenc}

\newcommand{\Caltech}{Department of Astronomy, California Institute of Technology, Pasadena, CA 91125, USA}

\newcommand{\Macquarie}{School of Mathematical and Physical Sciences, Macquarie University, Balaclava Road, North Ryde, NSW 2109, Australia}

\newcommand{\JPL}{Jet Propulsion Laboratory, California Institute of Technology, 4800 Oak Grove Drive, Pasadena, CA 91109, USA}
\newcommand{\MITEAPS}{Department of Earth, Atmospheric, and Planetary Sciences, Massachusetts Institute of Technology, Cambridge, MA 02139, USA}
\newcommand{\MITKavli}{Department of Physics and Kavli Institute for Astrophysics and Space Research, Massachusetts Institute of Technology, Cambridge, MA 02139, USA}
\newcommand{\UCI}{Department of Physics \& Astronomy, The University of California, Irvine, Irvine, CA 92697, USA}
\newcommand{\Carnegie}{Earth and Planets Laboratory, Carnegie Institution for Science, 5241 Broad Branch Road, NW, Washington, DC 20015, USA}

\newcommand{\Princeton}{Department of Astrophysical Sciences, Princeton University, 4 Ivy Lane, Princeton, NJ 08540, USA}

\newcommand{\Tsinghua}{Department of Astronomy, Tsinghua University, Beijing 100084, China}

\newcommand{\UCO}{UC Observatories, University of California, Santa Cruz, CA 95064, USA}

\newcommand{\WMKO}{W.\ M.\ Keck Observatory, 65-1120 Mamalahoa Hwy, Kamuela, HI 96743, USA}
\newcommand{\SSL}{Space Sciences Laboratory, University of California, Berkeley, CA 94720, USA}

\newcommand{\UCB}{Department of Astronomy, 501 Campbell Hall, University of California, Berkeley, CA 94720, USA}
\newcommand{\UCLA}{Department of Physics \& Astronomy, University of California Los Angeles, Los Angeles, CA 90095, USA}
\newcommand{\nexsci}{NASA Exoplanet Science Institute/Caltech-IPAC, California Institute of Technology, Pasadena, CA
91125, USA}
\newcommand{\COO}{Caltech Optical Observatories, California Institute of Technology, Pasadena, CA 91125, USA}

\newcommand{\Schmidt}{Astrophysics \& Space Institute, Schmidt Sciences, New York, NY 10011, USA}

\newcommand{\ND}{Department of Physics and Astronomy, University of Notre Dame, Notre Dame, IN 46556, USA}
\newcommand{\Geneva}{Observatoire Astronomique de l'Université de Genève, Chemin Pegasi 51, 1290 Versoix, Switzerland}
\newcommand{\kansas}{Department of Physics \& Astronomy, University of Kansas, 1082 Malott,1251 Wescoe Hall Dr., Lawrence, KS 66045, USA}

\newcommand{\DGPSCaltech}{Division of Geological and Planetary Sciences,
1200 E California Blvd, Pasadena, CA, 91125, USA}
\newcommand{\GSFC}{NASA's Goddard Space Flight Center, Greenbelt, MD 20771, USA}
\newcommand{\IFA}{Institute for Astronomy, University of Hawai`i, 2680 Woodlawn Drive, Honolulu, HI 96822, USA}

\usepackage{lineno}

\newcommand{\e}{$_\oplus$}
\newcommand{\p}{$\pm$}
\newcommand{\cms}{cm~s$^{-1}$}
\newcommand{\teff}{T$_{\rm eff}$}
\newcommand{\gaia}{\textit{Gaia}}
\newcommand{\rs}{$R_\star$}

\newcommand{\sol}{$_\odot$}
\newcommand{\ror}{$R_{\rm p}/R_\star$}

\begin{document}

\title{TOI-6324b: An Earth-Mass Ultra-Short-Period Planet Transiting a Nearby M Dwarf}

\correspondingauthor{Rena A. Lee}
\email{renaalee@hawaii.edu}

\author[0000-0001-7058-4134]{Rena A. Lee}
\altaffiliation{NSF Graduate Research Fellow}
\affiliation{\IFA}

\author[0000-0002-8958-0683]{Fei Dai}
\affiliation{\IFA}
\affiliation{\DGPSCaltech}
\affiliation{\Caltech}

\author[0000-0001-8638-0320]{Andrew W. Howard}
\affiliation{\Caltech}

\author[0000-0003-1312-9391]{Samuel Halverson}
\affil{\JPL}

\author[0000-0002-0672-9658]{Jonathan Gomez Barrientos}
\affiliation{\DGPSCaltech}

\author[0000-0002-0371-1647]{Michael Greklek-McKeon}
\affiliation{\DGPSCaltech}

\author[0000-0002-5375-4725]{Heather A. Knutson}
\affiliation{\DGPSCaltech}

\author[0000-0003-3504-5316]{Benjamin J.\ Fulton}
\affiliation{\Caltech}

\author[0000-0001-7409-5688]{Guðmundur Stef\'ansson}
\affiliation{Anton Pannekoek Institute for Astronomy, University of Amsterdam, Science Park 904, 1098 XH Amsterdam, The Netherlands}

\author[0000-0001-8342-7736]{Jack Lubin}
\affiliation{\UCLA}

\author[0000-0002-0531-1073]{Howard Isaacson}
\affiliation{\UCB}
\affiliation{Centre for Astrophysics, University of Southern Queensland, Toowoomba, QLD, Australia}

\author[0000-0002-4480-310X]{Casey L. Brinkman}
\altaffiliation{NSF Graduate Research Fellow}
\affiliation{\IFA}

\author[0000-0003-2657-3889]{Nicholas Saunders}
\altaffiliation{NSF Graduate Research Fellow}
\affiliation{\IFA}

\author[0000-0003-3244-5357]{Daniel Hey}
\affiliation{\IFA}

\author[0000-0001-8832-4488]{Daniel Huber}
\affiliation{\IFA}

\author[0000-0002-3725-3058]{Lauren M. Weiss}
\affiliation{\ND}

\author[0000-0003-0638-3455]{Leslie A. Rogers}
\affiliation{Department of Astronomy \& Astrophysics, University of Chicago, 5640 S Ellis Avenue, Chicago, IL 60637, USA}

\author[0000-0003-3993-4030]{Diana Valencia}
\affiliation{Centre for Planetary Sciences, University of Toronto, 1265 Military Trail, Toronto, ON, M1C 1A4, Canada} 

\author[0000-0002-9479-2744]{Mykhaylo Plotnykov}
\affiliation{Department of Physics, University of Toronto, Toronto, ON M5S 3H4, Canada}

\author[0000-0003-0062-1168]{Kimberly Paragas}
\affiliation{\DGPSCaltech}

\author[0000-0003-2215-8485]{Renyu Hu}
\affiliation{\JPL}
\affiliation{\DGPSCaltech}

\author[0000-0002-7127-7643]{Te Han}
\affil{\UCI}

\author[0000-0003-0967-2893]{Erik A. Petigura}
\affiliation{\UCLA}

\author[0000-0003-3856-3143]{Ryan Rubenzahl}
\affiliation{Center for Computational Astrophysics, Flatiron Institute, 162 Fifth Avenue, New York, NY 10010, USA}

\author[0000-0002-5741-3047]{David~R.~Ciardi}
\affil{\nexsci}

\author[0000-0002-5812-3236]{Aaron Householder}
\affiliation{\MITEAPS}
\affiliation{\MITKavli}

\author[0000-0003-0742-1660]{Gregory J. Gilbert}
\affil{\UCLA}

\author[0000-0001-7664-648X]{J.~M.~Joel Ong}
\altaffiliation{NASA Hubble Fellow}
\affiliation{\IFA}

\author[0000-0002-2696-2406]{Jingwen Zhang}
\altaffiliation{NASA FINESST Fellow}
\affil{\IFA}

\author{Jacob Luhn}
\affiliation{\JPL}

\author{Luke Handley}
\affil{\Caltech}

\author[0000-0001-7708-2364]{Corey Beard}
\altaffiliation{NASA FINESST Fellow}
\affiliation{\UCI}

\author[0000-0002-8965-3969]{Steven Giacalone}
\affiliation{\Caltech}

\author[0000-0002-5034-9476]{Rae Holcomb}
\affil{\UCI}

\author[0000-0002-4290-6826]{Judah Van Zandt}
\affil{\UCLA}

\author[0000-0002-6525-7013]{Ashley D.\ Baker}
\affiliation{\COO}



\author[0009-0008-9808-0411]{Max Brodheim}
\affiliation{\WMKO}

\author{Matt Brown}
\affiliation{\WMKO}


\author[0000-0002-9003-484X]{David Charbonneau}
\affil{Center for Astrophysics, Harvard \& Smithsonian, 60 Garden Street, Cambridge, MA 02138, USA}

\author[0000-0001-6588-9574]{Karen A. Collins}
\affil{Center for Astrophysics, Harvard \& Smithsonian, 60 Garden Street, Cambridge, MA 02138, USA}

\author[0000-0002-1835-1891]{Ian J. M. Crossfield}
\affiliation{\kansas}

\author[0009-0000-3624-1330]{William Deich}
\affil{\UCO}


\author[0000-0002-9332-2011]{Xavier Dumusque}
\affiliation{\Geneva}


\author[0009-0004-4454-6053]{Steven R.\ Gibson}
\affiliation{\COO}

\author[0000-0002-0388-8004]{Emily Gilbert}
\affiliation{\JPL}




\author[0000-0002-7648-9119]{Grant M.\ Hill}
\affiliation{\WMKO}

\author[0000-0002-6153-3076]{Bradford Holden}
\affil{\UCO}



\author[0000-0002-4715-9460]{Jon M. Jenkins}
\affiliation{NASA Ames Research Center, Moffett Field, CA 94035, USA}


\author{Stephen Kaye}
\affiliation{\COO}

\author[0000-0003-2451-5482]{Russ R. Laher}
\affiliation{\nexsci}

\author[0009-0004-0592-1850]{Kyle Lanclos}
\affiliation{\WMKO}

\author[0000-0002-1422-4430]{W. Garett Levine}
\affiliation{Department of Astronomy, Yale University, New Haven, CT 06511, USA}





\author[0009-0008-4293-0341]{Joel Payne}
\affiliation{\WMKO}

\author[0000-0001-7047-8681]{Alex S. Polanski}
\affil{\kansas}
\affil{Lowell Observatory, 1400 W Mars Hill Rd, Flagstaff, AZ 86001, USA}

\author[0000-0002-7893-1054]{John O'Meara}
\affiliation{\WMKO}


\author[0000-0003-2058-6662]{George R. Ricker}
\affiliation{\MITKavli}

\author{Kodi Rider}
\affiliation{\SSL}

\author[0000-0003-0149-9678]{Paul Robertson}
\affil{\UCI}


\author[0000-0001-8127-5775]{Arpita Roy}
\affiliation{\Schmidt}


\author[0000-0001-5347-7062]{Joshua E. Schlieder}
\affiliation{\GSFC}

\author[0000-0002-4046-987X]{Christian Schwab}
\affil{\Macquarie}

\author[0000-0002-6892-6948]{Sara Seager}
\affiliation{\MITEAPS}
\affiliation{\MITKavli}
\affil{Department of Aeronautics and Astronautics, Massachusetts Institute of Technology, Cambridge, MA 02139, USA}

\author[0000-0003-3133-6837]{Abby P.\ Shaum}
\affiliation{\Caltech}

\author[0009-0007-8555-8060]{Martin M.\ Sirk}
\affiliation{\SSL}


\author[0009-0008-5145-0446]{Stephanie Striegel}
\affil{SETI Institute, Mountain View, CA 94043 USA}
\affil{NASA Ames Research Center, Moffett Field, CA 94035, USA}

\author[0009-0008-2801-5040]{Johanna Teske}
\affil{\Carnegie}


\author{John Valliant}
\affiliation{\WMKO}


\author[0000-0001-6763-6562]{Roland Vanderspek}
\affiliation{\MITKavli}

\author[0000-0002-1871-6264]{Gautam Vasisht}
\affil{\JPL}

\author[0000-0002-6092-8295]{Josh Walawender}
\affiliation{\WMKO}


\author[0000-0002-6937-9034]{Sharon Xuesong Wang}
\affil{\Tsinghua}

\author[0000-0002-4265-047X]{Joshua N. Winn}
\affil{\Princeton}

\author{Edward Wishnow}
\affiliation{\SSL}



\author[0000-0002-4037-3114]{Sherry Yeh}
\affiliation{\WMKO}

\begin{abstract}
     We report the confirmation of TOI-6324 b, an Earth-sized (1.059 \p\ 0.041 R\e) ultra-short-period (USP) planet orbiting a nearby ($\sim$20 pc) M dwarf. Using the newly commissioned Keck Planet Finder (KPF) spectrograph, we have measured the mass of TOI-6324 b  1.17 \p\ 0.22 M\e. Because of its extremely short orbit of just $\sim$6.7 hours, TOI-6324 b is intensely irradiated by its M dwarf host, and is expected to be stripped of any thick, H/He envelope. We were able to constrain its interior composition and found an iron core mass fraction (CMF = 27$\pm$37\%) consistent with that of Earth ($\sim$33\%) and other confirmed USPs. TOI-6324 b is the closest to Earth-sized USP confirmed to date. TOI-6324 b is a promising target for JWST phase curve and secondary eclipse observations (Emission Spectroscopy Metric = 25) which may reveal its surface mineralogy, day-night temperature contrast, and possible tidal deformation. From 7 sectors of TESS data, we report a tentative detection of the optical phase curve variation with an amplitude of $42\pm28$ ppm.  
\end{abstract}

\section{Introduction} \label{sec:intro}
Detecting and characterizing Earth-like planets is a leading pursuit in the field of exoplanetary astronomy. Photometric surveys, such as the Transiting Exoplanet Survey Satellite (TESS) mission \citep{ricker2015}, have enabled the identification of hundreds of Earth-sized planets and planet candidates. To begin to explore the interiors of these distant worlds, we require precise mass and radius measurements. However, true Earth analogs, with radial velocity (RV) semi-amplitudes on the order of $\sim$10 \cms, are not yet routinely detectable even with the current state-of-the-art instrumentation. The best insight into terrestrial planet composition, therefore, is currently offered by Earth-sized ultra-short-period planets \citep[USPs;][]{sanchis2014} whose RV semi-amplitudes are typically an order of magnitude larger than their longer-period counterparts. 

USPs (R$_{\rm p}$ $<$ 2R\e, P$_{\rm orb}\,<$ 1 d) are expected to be stripped rocky cores \citep{sanchis2014,lundkvist_hot_2016}, since they are so intensely irradiated by their host stars \citep[$>$ 650 S\e\ around Sun-like stars;][]{lundkvist_hot_2016}. This has been observationally confirmed via transmission and emission spectroscopy for several USPs \citep[e.g., GJ 1252 b, LHS 3844 b and GJ 367 b,][respectively]{Crossfield2022,kreidberg_absence_2019,zhang_gj_2024}. Precise mass and radius measurements of USPs therefore allow us to directly probe their compositions, with fewer unknown parameters than planets that may have a gaseous atmosphere. Still, non-negligible volatile envelopes have been suggested to be present on larger ($\sim$1.5-2 R\e) USPs, such as TOI-561 b \citep{brinkman_toi-561_2023} and 55 Cnc e \citep{demory_map_2016,Tsiaras2016,dai_homogeneous_2019}, though there may be other reasons for their anomolously low densities. Earth-sized and smaller USPs may be less likely to retain secondary atmospheres because of their lower escape velocities. Precise composition constraints for such planets are paramount for informing the formation pathways of terrestrial planets. For example, confirming a more complete sample of Earth-sized and smaller exoplanets may verify the mass-dependence of compositional outcomes during the final stages of the giant impact phase of planet formation \citep[e.g.,][]{scora2022}. 

In addition to being ideal for precise interior composition studies, USPs are especially promising targets for detailed observations of their surface characteristics with the James Webb Space Telescope (JWST). TOI-6324 b, confirmed in the present work, has one of the highest emission spectroscopy metrics \citep[ESM = 25;][]{kempton2018}, which is a proxy for the expected signal-to-noise of secondary eclipse measurements by JWST. It is one of the most similar in size to Earth, even among previously-selected JWST targets. Measuring the phase offset, longitudinal temperature distribution (day-night temperature contrast), and albedo from the phase curve will reveal the presence or absence of a secondary atmosphere \citep[e.g.,][]{demory_map_2016,angelo2017,kreidberg_absence_2019,Hu2024,mansfield2024}. Furthermore, USPs may be subject to significant deformation due to tidal interactions \citep{Dai2024}, and the phase curve may constrain the degree of tidal distortion \citep{Price2020}. Secondary eclipse emission spectra, in the absence of a secondary atmosphere, will constrain the dominant surface rock type of the planet \citep{hu2012,kreidberg_absence_2019,whittaker2022,zhang_gj_2024}. 

In this work, we set out to confirm and further characterize TOI-6324 b. This paper is organized as follows: we detail the properties of the host star TOI-6324 (TIC 372207328) in Sec.\ \ref{sec:stellar}. We present our transit modeling, phase curve modeling, and radius measurement of TOI-6324 b  in Sec.\ \ref{sec:phot}, and our RV monitoring and mass measurement in Sec.\ \ref{sec:spec}. We discuss the interior composition and possible tidal deformation of TOI-6324 b in Sec.\ \ref{sec:discussion}, along with prospective JWST observations. We conclude with a summary of our key findings in Sec.\ \ref{sec:sum}. 

\section{Host Star Properties}\label{sec:stellar}
\subsection{Spectroscopic Analysis}\label{ssec:speca}
A spectrum of TOI-6324 (TIC 372207328; 2MASS J22032128+6729596; Gaia DR2 2220012421430629632) was obtained with the High Resolution Echelle Spectrometer \citep[HIRES;][]{Vogt2014} on the 10-m Keck I Telescope situated at Mauna a Wākea (Maunakea). The observation was taken on 2023-11-26UT at 900 s exposure time, with a peak SNR of $\sim$150. We employed the {\tt SpecMatch-Emp} routine \citep{Yee2017} to empirically derive spectroscopic parameters by cross-matching the observed HIRES spectrum with an extensive library of well-calibrated stellar spectra observed by the California Planet Search collaboration. This routine mitigates the significant systematic effects found in direct spectral modeling of low-mass stars, such as poor continuum identification and normalization, or incomplete molecular line lists, which are especially troublesome for cool stars. From {\tt SpecMatch-Emp} we find \teff\ = 3247 \p\ 70 K, consistent with the TESS Input Catalog estimate \citep[TIC;][]{TIC,muirhead2018}. We also find [Fe/H] = -0.32 \p\ 0.09, consistent with the estimate from \citet{dittman2016}, and an upper limit on $v\sin i$ of < 2 km s$^{-1}$.

We utilized the Python package {\tt isoclassify} \citep{huber2017} to derive stellar parameters based on our derived spectroscopic stellar parameters and \gaia\ astrometry. {\tt Isoclassify} directly computes the stellar radius (\rs) using the Stefan-Boltzmann Law, based on the observed \textit{K}-band magnitude, \gaia\ parallax, and \teff. We adopted the MESA Isochrones and Stellar Tracks \citep[MIST;][]{MIST} and adhered to the default settings recommended by {\tt isoclassify}. We compared our results with the empirical M dwarf mass-luminosity relation from \citet{mann2019}, and found that they are consistent within model uncertainties. The derived stellar parameters are given in the top section of Table \ref{tab:params}.

\subsection{Additional Companion Search}
 To validate transiting exoplanets and assess possible contamination of bound or unbound companions on the derived planetary radii \citep{ciardi2015}, we observed TOI-6324 using near-infrared adaptive optics (AO) imaging and optical speckle imaging. The AO observations were conducted on 2023-08-05UT using NIRC2 on Keck-II as part of the K2 and TESS follow-up and validation survey by \citet{Schlieder2021}. We utilized a 3-point dither pattern. The narrow-angle mode provided a field of view of ~10\arcsec\ with a pixel scale of $\sim0.0099442\arcsec$/pixel, using the narrow-band K$_{cont}$ filter $(\lambda_o = 2.2706; \Delta\lambda = 0.0296~\mu$m).

 \begin{figure}[!h]
    \centering
    \includegraphics[width=8cm]{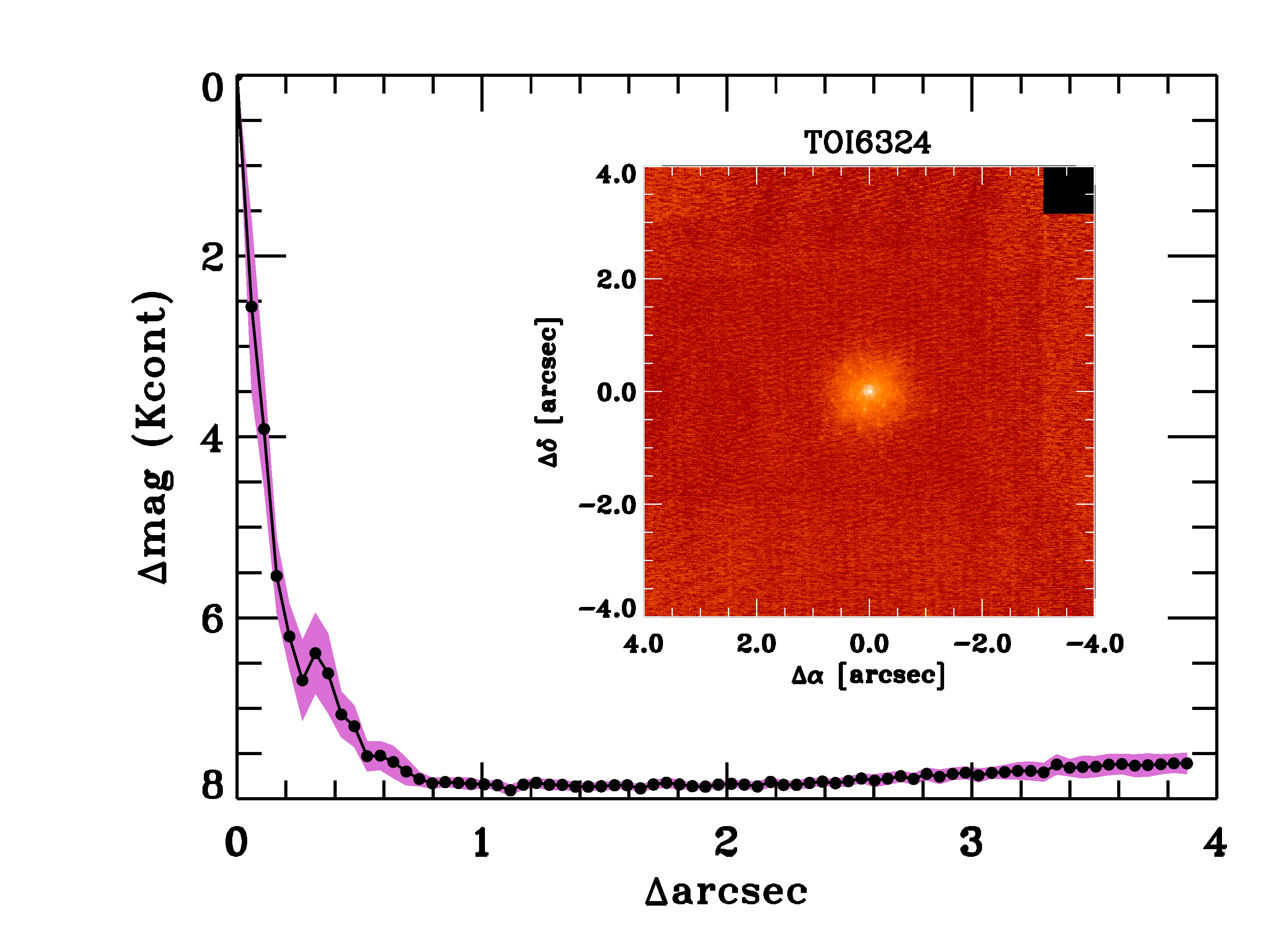}
    \caption{NIR AO imaging and sensitivity curve. The black points represent the 5$\sigma$ detection limit and the purple region represents the 1$\sigma$ azimuthal dispersion of the contrast determination. \textit{Inset:} AO image of TOI-6324 showing no evidence for a close-in companion.}
    \label{fig:aoimaging}
\end{figure}
 
 Each frame had an integration time of 2.9 seconds, totaling 26.1 seconds on-source. Flat fields were taken on-sky, dark subtracted, and median averaged, and sky frames were created from the median average of dithered science frames. These were sky-subtracted and flat-fielded, resulting in a combined image with a resolution of 0.053\arcsec. Sensitivity was assessed by injecting simulated sources around the primary target at 20$^\circ$ intervals, scaling brightness until detection at 5$\sigma$ significance \citep{furlan2017}. We ruled out $\delta$mag $<$7 companions within 0.5\arcsec\ (Fig.\ \ref{fig:aoimaging}).

In addition to high contrast imaging, we investigated \gaia\ DR3 to search for stellar companions to TOI-6324. We identified no co-moving sources in \gaia\ DR3 within a radius of 10\arcsec. We further assessed the astrometry of TOI-6324 for evidence of a stellar companion. The \gaia\ DR3 Renormalised Unit Weight Error (RUWE) value represents the reduced $\chi^2$ of \gaia's single-star astrometric solution and can be used to identify unresolved binaries \citep{Belokurov2020} at separations of $\sim$0.1-1.0\arcsec\ \citep{Wood2021}. For TOI-6324, RUWE = 1.165, well below the empirical threshold (RUWE = 1.4) at which binarity is likely.

\subsection{Age} \label{ssec:age}
It is notoriously difficult to determine the age of M dwarfs using the methods applied to more massive stars. Their asteroseismic oscillations are too small to be detected with current capabilities \citep{rodriguez2016,berdinas2017}, and their isochrone ages are poorly calibrated \citep{engle2018}. To assess the age of TOI-6324, we turned to gyrochronology \citep{Barnes2003,Barnes2007}, which relates the rotation rate of a star to its age based on empirically-calibrated, mass-dependent spin-down rates \citep{Skumanich1972,Barnes2007}. However, we did not recover a robust rotation signal from a Lomb-Scargle periodogram \citep{Lomb1976,Scargle1982} of TESS \citep[][]{ricker2015} light curves (further described in Sec.\ \ref{sec:phot}) using a standard {\tt lightkurve} routine \citep{lightkurve}. This may be attributed to the approximate upper limit of $P=13.7$ days imposed by the length of the TESS orbit. Older stars with rotation periods longer than 13.7 days are difficult to age-date using TESS photometry alone. \citet{newton2016} also report a non-detection of the rotation period of TOI-6324, utilizing long-term photometric monitoring data from MEarth \citep{berta2012,irwin2015}. Furthermore, using {\tt BANYAN $\Sigma$} \citep{gagne2018}, we found the \gaia\ DR3 kinematics of TOI-6324 are consistent with that of field stars. It is therefore unlikely that TOI-6324 is a member of any known nearby young moving groups. We are unable to derive a precise age estimate based on gyrochronology or by association, but it is likely that TOI-6324 is not a young ($\lesssim$100 Myr) star. 

We calculated the probability that TOI-6324 is a member of the thick disk using the prescription of \citet{Bensby}. TOI-6324 has a $TD/D=84$, i.e., it is likely a thick disk member based on its Galactic UVW velocity (U, V, W = 9.4$\pm$1.0, -65.7$\pm$2.1, -60.4$\pm0.5$ km~s$^{-1}$) after correcting for the Local Standard of Rest (LSR). This is consistent with the trend that USPs tend to occur around older stars \citep{Schmidt2024}. We adopted the LSR as U$_\odot$,V$_\odot$,W$_\odot$ = 10.0$\pm1.0$, 11.0$\pm2.0$, 7.0$\pm0.5$ km~s$^{-1}$ from \citet{Bland-Hawthorn}. Unfortunately, we do not have $[\alpha/Fe]$ data to back up the thick disk membership using the host star’s abundances, though we do find a non-detection of H$\alpha$ emission in the HIRES spectrum (Sec.\ \ref{ssec:speca}).

\begin{deluxetable*}{lccc}
\tablecaption{TOI-6324 system parameters. For parameters derived in this work, the \textit{Value} column reports median posterior values and 68.3\% confidence interval.} 
\label{tab:params}
\tablehead{
\colhead{Parameter} & \colhead{Symbol} & \colhead{Prior} & \colhead{Posterior (Median and 68.1\% CI)}}
\startdata
\textbf{Stellar}\\
TIC ID $^\dagger$  & & & 372207328  \\
R.A. $^\dagger$  & $\alpha$ & & 22:03:22.69\\
Dec. $^\dagger$  & $\delta$ & & +67:29:55.2\\ 
$V$ (mag) $^\dagger$ &  & &	$13.386\pm0.026$ \\
$K_s$ (mag) $^\dagger$ & & & $8.691\pm0.022$\\
Effective Temperature (K)& $T_{\text{eff}}$ & &$3247\pm70$ \\
Surface Gravity$~(\text{cm~s}^{-2})$ & $\log~g$ & & $4.93 \pm  0.006$\\
Iron Abundance $~(\text{dex})$ & $[\text{Fe/H}]$ & & $-0.32 \pm 0.09$ \\
Rotational Broadening $~(\text{km~s}^{-1}$) & $v~\text{sin}~i$ &  & $<$2\\
Stellar Mass$~(M_{\odot})$ & $M_{\star} $ & & $0.269 \pm 0.012$ \\
Stellar Radius$~(R_{\odot})$ & $R_{\star}$ & & $0.293 \pm 0.010$\\
Stellar Density (g cm$^{-3}$) & $\rho_\star$  & & $10.7 \pm 1.1$\\
Limb Darkening q$_1$ \citep{exoplanet:kipping13} &  & $U$ (0, 1) & $0.40 \pm0.28$ \\
Limb Darkening q$_2$ \citep{exoplanet:kipping13} &   & $U$ (0, 1) & 0.13$^{+0.32}_{-0.13}$\\
Parallax (mas) $^\ddagger$ & $\pi$ & & $48.6414 \pm 0.0146$\\
Distance (pc) $^\ddagger$ & \textit{d} &  & $20.5586 \pm 0.0061$ \\
\hline
\textbf{Planetary}\\
Emission Spectroscopy Metric & ESM & & 25\\
Equilibrium Temperature$^*$ & $T_{\text{eq}}$ & & $1216\pm60$ \\
Planet/Star Radius Ratio & \ror\  & $\mathcal{N}\,\propto$ transit depth & $0.0331\pm0.0006$  \\
Time of Conjunction (BJD-2457000) & $T_c$  &  $\mathcal{N}$ (1738.5829, 10) & $1739.97928\pm0.00068$  \\
Impact Parameter & $b$  & $U$ (0, 1+R$_{\rm p}$/R$_{\star}$) & $0.75 \pm 0.03$\\
Scaled Semi-major Axis & $a/R_\star$  & & $3.17\pm0.20$  \\
Orbital Inclination (deg) & $i$  & & $74.6 \pm 0.7$ \\
Orbital Eccentricity  & $e$  & 0 (fixed)& 0  \\
Orbital Period (days) & $P_{\rm orb}$   & $\mathcal{N}$ (0.279211, 10) & $0.2792210\pm0.00000010$ \\
Planetary Radius ($R_\oplus$)  & $R_{\rm p}$ & & $1.059 \pm 0.041$ \\
RV Semi-amplitude (m~s$^{-1}$)  & $K$ & Jeffreys (0.1, 30)  & $2.69\pm0.51$ \\
Planetary Mass ($M_\oplus$)  & $M_{\rm p}$ & & $1.17 \pm 0.22$\\
\hline
KPF RV Jitter (m~s$^{-1}$)  & $\sigma_{\rm jit,KPF}$ & Jeffreys (0.1, 10) & $1.31\pm0.33$\\
GP Kernel Amplitude  (m~s$^{-1}$) & \textit{h} & Jeffreys (0.1, 100) & 21.44$^{+3.66}_{-2.81}$ \\
GP Length (days) & $l$ & Jeffreys (2P, $\sqrt{2}\tau$) &  1.76 $^{+0.29}_{-0.40}$\\
\enddata
\tablecomments{$^\dagger$TICv8; $^\ddagger$\textit{Gaia} DR3; $^*T_{\text{eq}}$ assumes a low albedo of A$_B$=0.1}
\end{deluxetable*}

\section{Photometric Analysis}\label{sec:phot}

\subsection{Observations \& Data}\label{ssec:obs_p}
TOI-6324 was observed by TESS during Sectors 16, 17, 18, 24, 58, 77, and 78 (September-November 2019, April-May 2020, November 2022, and April-May 2024). The data were processed by the TESS Science Processing Operations Center \citep[SPOC at NASA Ames Research Center;][]{jenkins_tess_2016} and the MIT Quick-Look Pipeline \citep[QLP;][]{Huang2020}. The transit signal of TOI-6324 b at 0.279 days was first alerted in a faint-star search in QLP, reported on the Exoplanet Follow-up Observing Program (ExoFOP
\footnote{\url{https://exofop.ipac.caltech.edu/tess/target.php?id=372207328}}), and the SPOC pipeline detected the transit signature in every sector observed. We collected the 2-minute cadence SPOC light curves using the Python package {\tt lightkurve} \citep{lightkurve}. We applied transit-depth bias correction to the Year 2 data (Sectors 16, 17, 18, and 24) which were affected by a sky background bias issue documented in the S27 Data Release Notes. We utilized the Presearch Data Conditioning Simple Aperture Photometry (PDC-SAP) light curves \citep{Smith2012,Stumpe2012,Stumpe2014,twicken:PA2010SPIE} for transit modeling after removing data points with non-zero data quality flags.  

\subsection{Transit Modeling}\label{ssec:transit}
To prepare the PDC-SAP light curve for transit modeling we first removed instrumental and long-term stellar systematics using the {\tt wōtan} Python package \citep{hippke_wotan_2019}. We first masked the transit signal of TOI-6324 b based on the QLP transit parameters. We then used an iterative sigma-clipping spline fit ({\tt rspline}) with a width of 0.5 days to detrend the transit-masked light curve. 

Our transit model was constructed using the {\tt exoplanet} package \citep{foreman-mackey_exoplanet_2021}, a gradient-based probabilistic inference toolkit for modeling time-series astronomical data. We used the transit parameters reported by QLP (depth, duration, orbital period, and epoch) as the basis for uninformative priors (either uniform or broad normal distributions) to initialize our model. This allowed for a thorough and unconstrained search of the parameter space. We additionally imposed Gaussian stellar radius and density priors derived in this work (see Sec.\ \ref{sec:stellar} and Table \ref{tab:params}). Limb darkening coefficients q$_1$ and q$_2$ were re-parameterized following the following the quadratic law formulation of \citet{exoplanet:kipping13}. We assumed a circular orbit ($e$ = 0) for model simplicity. Since the orbital period is extremely short, we expect this to be a valid assumption due to tidal circularization. The transit model further included the planet-to-star radius ratio \ror\ derived from transit depth, and the impact parameter $b$. The orbital inclination $i$ was derived from $b$. We fit all transits assuming a constant period, since we did not identify any significant transit-timing variations. 

\begin{figure}[!ht]
    \centering
    \includegraphics[width=8cm]{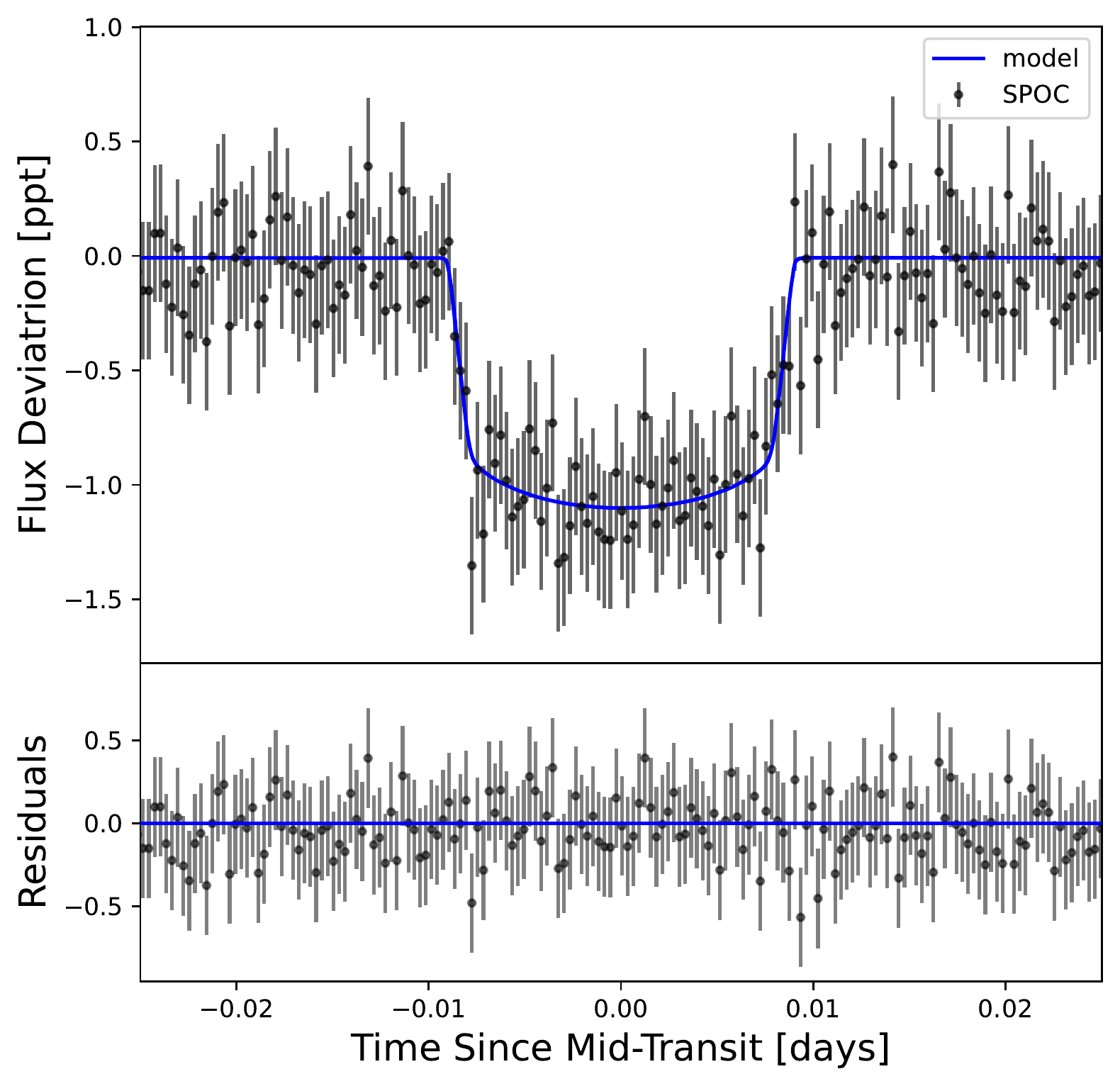}
    \caption{Binned and phase-folded SPOC PDC-SAP light curve of TOI-6324 revealing the transit signal of TOI-6324 b, with the best-fit transit model over-plotted in blue. The residuals (data minus model) are shown in the bottom panel.}
    \label{fig:lc}
\end{figure}

 A maximum a posteriori (MAP) transit model solution was determined from an initial non-linear optimzation using {\tt pyMC3} and set as the initial condition for sampling. We then explored the parameter space using No U-Turn Sampling \citep[NUTS;][]{NUTS} in a gradient-based Hamiltonian Monte Carlo \citep[HMC;][]{neal2011,betancourt2017} framework in {\tt pyCM3} to generate the transit parameter posteriors. We ran 2 chains for 10000 draws, with 5000 iterations to tune (15,000 steps total). We list the median values of the posterior distributions for the transit parameters in Table \ref{tab:params} along with the 68.3\% confidence intervals. Fig.\ \ref{fig:lc} shows the binned and phase-folded PDC-SAP light curve of TOI-6324, with a robust detection of TOI-6324 b, along with the best-fit transit model. We report a radius of $R_{\rm p}$ = 1.059 \p\ 0.041 R\e\ for TOI-6324 based on the best-fit planet-to-star radius ratio \ror\ = 0.0331 \p\ 0.0006 and our derived stellar radius of $R_{\star}$ = 0.293 \p\ 0.010 R\sol (Table \ref{tab:params}).

\subsection{Additional Photometry Sources}
In addition to the SPOC light curves, we obtained TESS-\textit{Gaia} light curves \citep[TGLC;][]{han_tessgaia_2023}, which are created from FFIs forward-modeled with the effective point-spread function (PSF) to remove contamination from nearby stars. Based on \gaia\ DR3 positions and magnitudes, these light curves achieve a very high-treatment account of dilution. We performed the identical analysis on the TGLC light curves, and find $R_{\rm p}$ = 1.075 \p\ 0.05 R\e\, consistent with the result from the SPOC light curves within 1$\sigma$. Furthermore, TOI-6324 b was observed with the 200-in Hale Telescope at Palomar Observatory for transit validation \citep{gomez2024}. The radius measurement of \citet{gomez2024}, using both ground-based and TESS SPOC light curves, is consistent with the measurement from this work. 

\subsection{Additional Transiting Planet Search}\label{ssec:addl}
No additional transiting planet candidates of TOI-6324 have been reported by QLP nor SPOC. To search for additional transit signals in the PDC-SAP light curves described in Sec.\ \ref{ssec:transit}, we masked the transits of TOI-6324 b and performed a Box-Least-Square \citep[BLS;][]{kovacs2002} analysis using the {\tt lightcurve.to\_periodogram} function of {\tt lightkurve} \citep{lightkurve}. We allowed for periods between 0.1 and 30 days, and transit durations between 0.1 and 2.15 hr. We phase-folded the light curve on the significant periodic signals from this search, and we did not identify any additional planetary transit signals. 

\subsection{Phase Curve}\label{ssec:pc}
TOI-6324 b is $\sim$20\% from the Roche limit, making it likely to be tidally distorted (see Sec.\ \ref{ssec:dist}). We searched the TESS light curves for evidence of phase curve variation and secondary eclipse signal from TOI-6324 b. The light curve was iteratively sigma-clipped and detrended in the same manner as in Sec.\ \ref{ssec:transit}, but with a wider spline width of $2*P_{orb}$ or $\sim$13.4 hr, to ensure any phase curve signal was not removed in the detrending step. The light curve was then phase-folded and binned to $\sim$10 minute intervals (50 bins total). We used {\tt batman} \citep{Kreidberg2015} to construct a secondary eclipse model based on the transit parameters derived in Sec.\ \ref{ssec:transit} (Table \ref{tab:params}), but with the limb darkening coefficients (u$_1$, u$_2$) set to 0. The eclipse time ($t_{sec}$) was fixed to a 0.5 phase offset from the primary transit time. 

In addition to the secondary eclipse, we modeled the out-of-eclipse phase curve variation, initially characterized by a combination of the illumination effect from reflected stellar light ($A_{ill}$), an offset of the peak illumination effect ($\theta$), and ellipsoidal light variation ($A_{ELV}$). $A_{ill}$ was fixed to be the same as the secondary eclipse depth ($\delta_{sec}$). We found that the TESS data do not have sufficient SNR to robustly constrain $\theta$, and thus we set $\theta=0$ for model simplicity. 

\begin{figure}[!ht]
\gridline{\fig{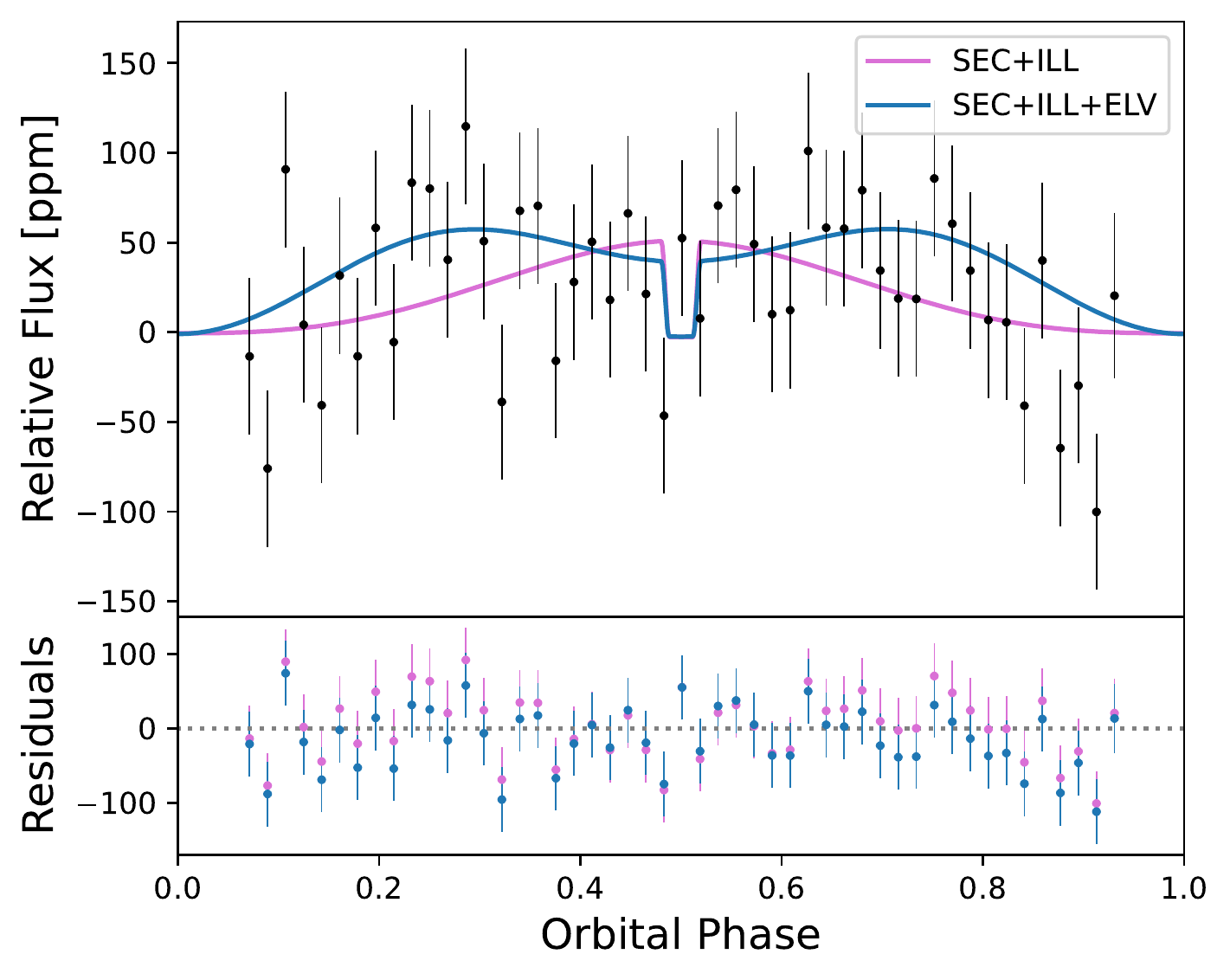}{0.6\textwidth}{\textbf{a)}}}
\gridline{\fig{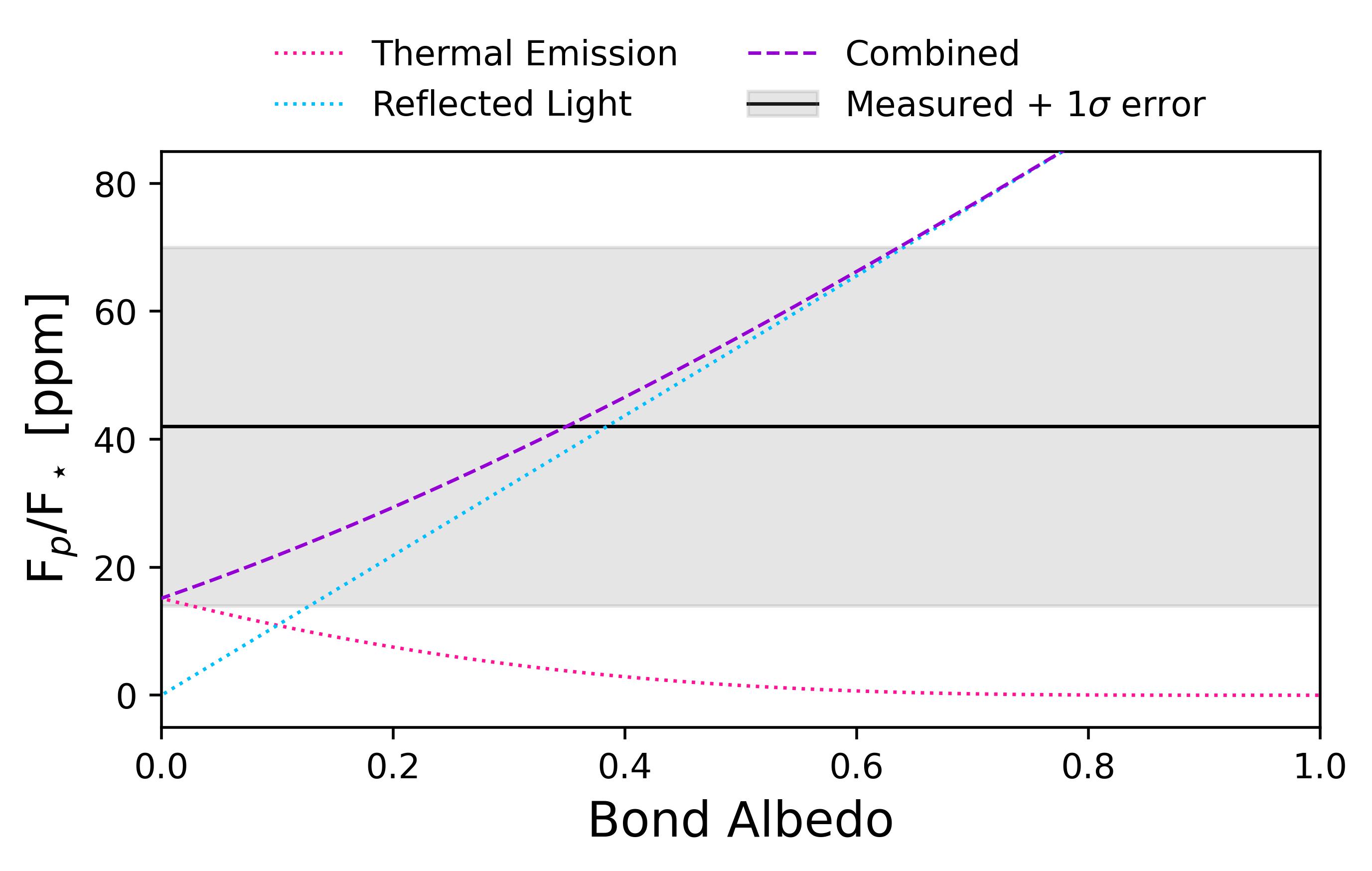}{0.6\textwidth}{\textbf{b)}}}
\caption{\textbf{a)}: Binned and phase-folded TESS light curve of TOI-6324. The pink curve shows the best-fit model without ellipsoidal light variation (ELV) in the phase curve model, and the blue curve shows the best-fit result which includes the ELV term. The residuals (data minus model) from each model are shown in the bottom panel. \textbf{b)}: Thermal emission (pink dotted line) and reflected light (blue dotted line) in the TESS band
as a function of the Bond Albedo. The flux ratio (F$_p$/F$_\star$) measured in this work is shown as the black line with the associated 1$\sigma$ confidence shaded in grey. The phase curve variation observed in the TESS band is likely attributed to a combination of both thermal emission and reflected stellar light (purple dashed line).}
\label{fig:pc}
\end{figure}

The ellipsoidal light variation $A_{ELV}$ can be due to both the tidal distortion of the star and the planet itself. We estimated that the stellar component is likely smaller than 1 ppm \citep[e.g.][]{Faigler}, however the planetary distortion can be substantial given the extremely short orbital period of the planet. We compared two phase curve models: with and without the ellipsoidal light variation $A_{ELV}$. Combined with the secondary eclipse model, the best-fit solutions were determined using the {\tt Levenberg-Marquardt} least-squares likelihood maximization method \citep[{\tt lmfit};][]{lmfit}. We employed a nested sampling routine using {\tt dynesty} \citep{Speagle2020,sergey_koposov_2024_12537467} to sample the posteriors and compare the Bayesian evidence ($\log(Z)$) of the two models, and found that the non-ELV model is preferred over the ELV-inclusive model ($\Delta\log(Z)=5$). We show in Fig.\ \ref{fig:pc}a the best-fit phase curve models with and without ELV. The posterior medians of the non-ELV and ELV models are consistent within $1\sigma$ uncertainties, and from the preferred non-ELV model we find $\delta_{sec}\equiv A_{ill}=42\pm28$ ppm ($<2\sigma$). From the ELV model, we report a tentative ($\sim1\sigma$) detection of $A_{ELV}=21\pm17$ ppm, corresponding to a Bond Albedo of $A_B=0.36^{+0.28}_{-0.36}$ (Fig.\ \ref{fig:pc}b).

\section{Spectroscopic Analysis}\label{sec:spec}
\subsection{Observations \& Data}\label{ssec:obs_s}
We obtained high-resolution spectra of TOI-6324 with the Keck Planet Finder \citep[KPF;][PI: A. Howard]{kpf} on Keck I. KPF is a high-resolution (R$\sim$98,000), optical (445-870 nm), fiber-fed echelle spectrometer newly commissioned in late 2022. The confirmation of TOI-6324 b in this work is among the first science results of KPF. KPF boasts an internal instrumental RV precision of $\sim$50 \cms\ and has demonstrated the capability to detect a 1.44 M\e\ USP to greater than 5$\sigma$ precision \citep[TOI-6255 b;][]{Dai2024}. We collected a total of 77 KPF spectra of TOI-6324 between 2023-06-30UT and 2024-01-01UT (see Table \ref{tab:rv}). Each exposure was 15 min and the average SNR achieved was 65 at 550 nm. We aimed to obtain multiple exposure per night in order to mitigate some of the instrumental drift. The spectra were reduced  using the publicly-accessible KPF Data Reduction Pipeline (DRP)\footnote{\url{https://github.com/Keck-DataReductionPipelines/KPF-Pipeline}}, the radial velocities were extracted with the template matching code \texttt{serval} \citep{SERVAL} after minor modifications for KPF \citep{Dai2024}. RVs from nights with only a single measurement, as well as nightly 3$\sigma$ outlier measurements, noted with a (*) in Table \ref{tab:rv}, were excluded, leaving 65 RV measurements taken over 23 nights for fitting.   

\startlongtable
\begin{deluxetable}{ccccc}
\tabletypesize{\scriptsize}
\tablecaption{Keck/KPF Radial Velocities of TOI-6324}
\label{tab:rv}
\tablehead{
\colhead{Time (BJD)} & \colhead{RV (m/s)} & \colhead{RV Uncertainty (m/s)} }
\startdata
2460126.076182692 $^\ast$ & -32.97 & 1.08 \\ 
2460130.010714815 $^\ast$& -48.53 & 1.25 \\ 
2460135.004611146 $^\ast$& -50.03 & 3.23 \\ 
2460135.099483773 $^\ast$& -37.17 & 1.70 \\ 
2460146.114391493 $^\ast$& -48.06 & 1.15 \\ 
2460150.082071852 $^\ast$& -20.9 & 1.12 \\ 
2460153.098518762 & -32.14 & 1.13 \\ 
2460153.127477107 & -37.35 & 1.23 \\ 
2460159.95573294 & -54.03 & 1.14 \\ 
2460159.987071331 & -52.14 & 1.13 \\ 
2460162.964025116 & -56.23 & 1.67 \\ 
2460163.075380926 & -54.04 & 1.63 \\ 
2460169.969600474 & -58.56 & 1.10 \\ 
2460170.080126852 & -52.53 & 1.16 \\ 
2460176.841209965 & -55.04 & 1.74 \\ 
2460177.049994664 & -53.08 & 1.72 \\ 
2460180.875071331 & -54.19 & 1.11 \\ 
2460181.011329491 & -51.37 & 1.13 \\ 
2460182.876525359 & -58.25 & 1.12 \\ 
2460183.013670394 & -59.47 & 1.09 \\ 
2460183.980496354 $^\ast$ & -48.57 & 1.10 \\ 
2460184.814127651 & -96.50 & 1.06 \\ 
2460184.914328611 & -92.69 & 1.05 \\ 
2460185.86687493 & -84.08 & 1.22 \\ 
2460185.935370845 & -81.48 & 1.13 \\ 
2460186.038032164 & -80.85 & 1.21 \\ 
2460187.792816146 & -51.63 & 1.14 \\ 
2460187.894425023 & -48.05 & 1.23 \\ 
2460188.019864653 & -47.90 & 1.26 \\ 
2460188.797937361 & -40.15 & 1.18 \\ 
2460188.825234838 & -40.28 & 1.23 \\ 
2460189.006935509 & -37.43 & 1.30 \\ 
2460189.962912512 & -38.08 & 1.17 \\ 
2460190.01121147 & -39.69 & 1.22 \\ 
2460190.05758022 $^\ast$ & -33.78 & 1.13 \\ 
2460192.812904292 $^\ast$ & -42.71 & 1.16 \\ 
2460196.778913785 $^\ast$ & -56.79 & 1.17 \\ 
2460196.813865821 & -49.94 & 1.15 \\ 
2460196.863170963 & -48.97 & 1.19 \\ 
2460197.766617256 $^\ast$ & -70.66 & 1.65 \\ 
2460197.793505319 & -45.34 & 1.31 \\ 
2460197.832056304 & -50.06 & 1.14 \\
2460198.802907853 & -34.68 & 1.05 \\ 
2460198.853179663 & -38.10 & 1.09 \\ 
2460198.877306925 & -38.51 & 1.03 \\ 
2460199.762083811 & -53.57 & 1.19 \\ 
2460199.801108737 & -53.61 & 1.15 \\ 
2460199.846550467 & -53.37 & 1.15 \\ 
2460200.760515219 & -45.52 & 1.13 \\ 
2460200.82617912 & -45.10 & 1.40 \\ 
2460200.856759512 & -49.55 & 1.38 \\ 
2460204.762007308 & -62.63 & 1.16 \\ 
2460204.793847212 & -65.26 & 1.15 \\ 
2460204.923963133 & -61.56 & 1.24 \\ 
2460208.770297217 & -64.04 & 1.11 \\ 
2460208.838113155 & -62.98 & 1.20 \\ 
2460208.867704152 $^\ast$ & -53.37 & 1.16 \\ 
2460209.752046775 & -50.23 & 1.41 \\ 
2460209.791566944 & -50.48 & 1.16 \\ 
2460209.82930193 & -53.22 & 1.23 \\ 
2460211.734217746 & -39.43 & 1.07 \\ 
2460211.84252483 & -45.36 & 1.13 \\ 
2460211.911891459 & -40.03 & 1.14 \\ 
2460252.886912332 & -24.69 & 1.12 \\ 
2460252.94776194 & -22.74 & 1.23 \\ 
2460273.721432364 & 19.40 & 0.96 \\ 
2460273.732485145 & 18.74 & 1.00 \\ 
2460273.746181982 & 17.70 & 1.02 \\ 
2460273.757201615 & 18.39 & 1.04 \\ 
2460273.768154097 & 16.20 & 1.06 \\ 
2460273.779157795 & 17.94 & 1.06 \\ 
2460273.79369893 & 15.90 & 1.02 \\ 
2460273.804679974 & 15.02 & 1.04 \\ 
2460273.815668433 & 15.01 & 1.00 \\ 
2460273.826684487 & 17.69 & 1.02 \\ 
2460310.761742466 & -88.81 & 1.09 \\ 
2460310.791932109 & -84.13 & 1.21 \\
\enddata
\tablecomments{$^\ast$Single-measurement nights and nightly 3$\sigma$ outlier measurements excluded from our RV fits.}
\end{deluxetable}

\subsection{Gaussian Process Model}\label{ssec:gp}
To disentangle the planetary RV modulation from the significant instrumental noise in the RV variations of TOI-6324, we employed a Gaussian Process (GP) regression model. We used the package {\tt radvel} \citep{fulton_radvel_2018} for this analysis. The correlated noise is likely a result of stochastic instrumental drift. Stellar pulsations which normally affect observations in more massive stars have very low amplitudes in M dwarfs, and because they occur on much shorter timescales, are averaged out in each exposure \citep{Kjeldsen1995,Chaplin2019}. In our analysis of the light curves, we did not identify any periodic variations that could be attributed to rotationally modulated stellar activity or pulsations. As such, we used a squared-exponential GP kernel in our model, rather than the widely-used quasi-periodic kernel \citep[e.g.,][]{grunblatt2015,rajpaul2015,brinkman2023}, for model simplicity \citep{dai_homogeneous_2019,dai_tks_2021,Dai2024}. This is included in the covariance matrix which is factored into the likelihood function. See \citet{Dai2024} for a detailed explanation of these expressions. 

The model basis included the orbital elements of orbital period $P_{\rm orb}$, time of conjunction $T_c$, and RV semi-amplitude $K$, and the eccentricity $e$ and argument of pericenter $\omega$ were included jointly as $\sqrt{e}\cos\omega$ and $\sqrt{e}\sin\omega$. We imposed Gaussian priors on P$_{\rm orb}$ and $T_c$ derived from our transit analysis (see Sec.\ \ref{sec:phot} and Table \ref{tab:params}). We placed a hard lower bound for semi-amplitude of $K > 0$, and we additionally imposed Jeffreys priors on the jitter term for KPF $\sigma_{\rm jit,KPF}$, as well as the GP hyperparameters of amplitude $h$ and length $l$. We set a lower bound of $l>1$ day to ensure the GP did not fit out the 6.7 hour planetary signal of TOI-6324 b. As in the transit model, we fixed the eccentricity to 0. 

\begin{figure*}
    \centering
    \includegraphics[width=16cm]{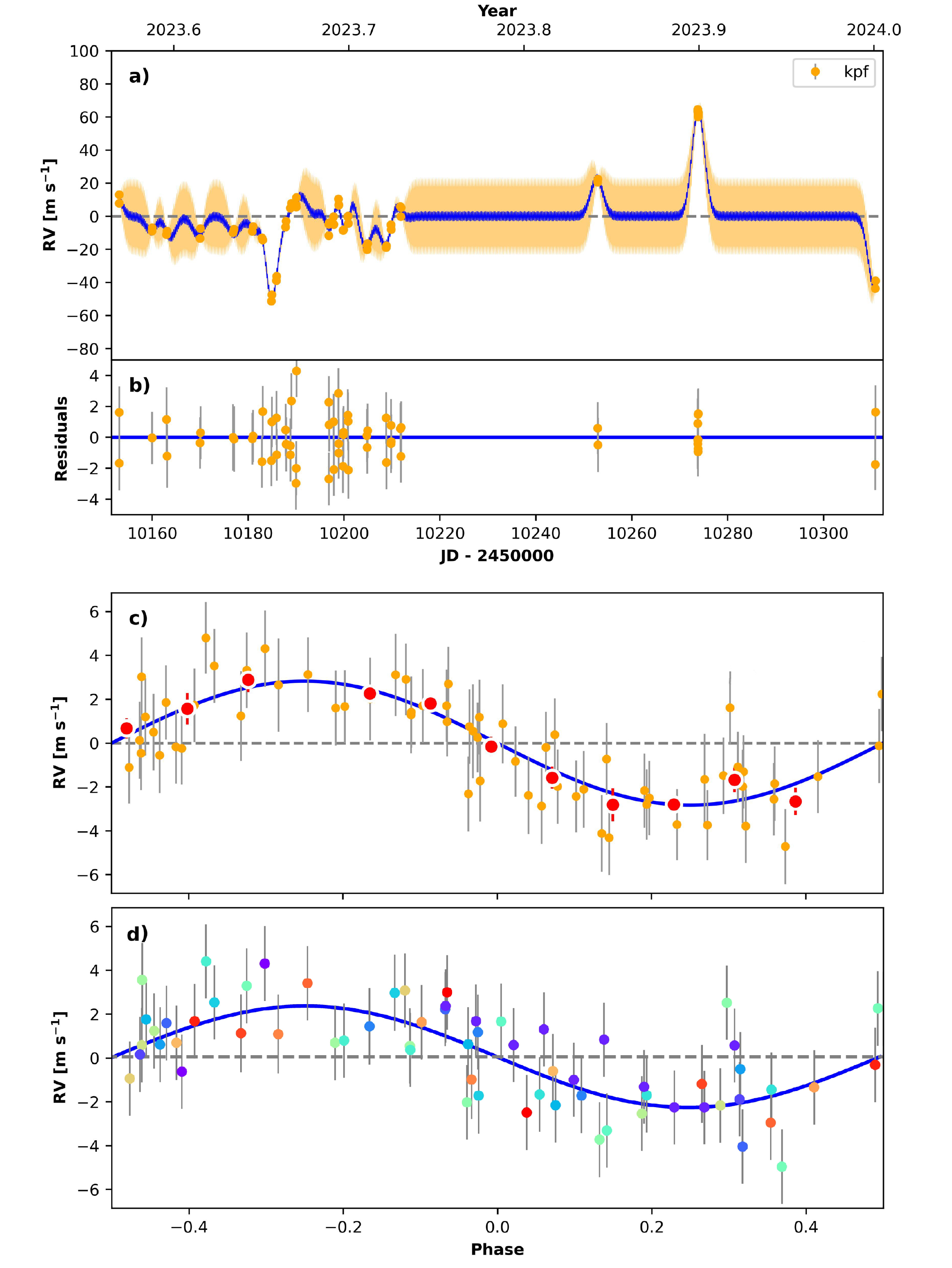}
    \caption{The radial velocity (RV) variations of TOI-6324 observed with KPF. \textbf{a)} time-series RV measurements along with the best-fit {\tt radvel} Gaussian Process (GP) model (blue line) used to remove stellar and instrumental systematics longer than the 6.7 hr orbital period of TOI-6324 b (see Sec.\ \ref{ssec:gp}). The yellow contour represents the uncertainty of the GP model. Data points have been median-offset by $-$49.94 m s$^{-1}$; the raw values are given in Table \ref{tab:rv}. \textbf{b)} residuals (data $-$ GP model). \textbf{c)} RV variations of TOI-6324 as a function of orbital phase, resulting from our GP model. The red points show RVs binned to 10\% of $P_{orb}$. \textbf{d)} RV variations of TOI-6324 as a function of orbital phase, resulting from our Floating Chunk Offset (FCO) analysis (see Sec.\ \ref{ssec:fco}). Data points are colored by observation date. The measured RV semi-amplitude from both the GP and FCO analyses are consistent within 1$\sigma$ (K$_{\rm GP}$ = 2.69 \p\ 0.51 m s$^{-1}$ and K$_{\rm FCO}$ = 2.19 \p\ 1.02 m s$^{-1}$)}.
    \label{fig:rv}
\end{figure*}

The standard {\tt radvel} routine employs {\tt emcee} \citep{emcee2013} to explore the parameter space in a Markov Chain Monte Carlo (MCMC) framework. We ran 128 walkers for 10$^4$ runs in 3 ensembles for parallelization, achieving a maximum Gelman-Rubin (Max G-R) statistic of 1.001.  We show the best-fit GP model in panel (a) and the best-fit Keplerian solution for TOI-6324 b in panel (c) of Fig.\ \ref{fig:rv}. The median values of the posterior distributions for the orbital parameters and GP hyperparameters are presented in Table \ref{tab:params}, respectively. We found that the RV semi-amplitude imposed by TOI-6324 b is robustly detected at 5$\sigma$ precision with $K=2.69 \pm 0.51\;{\rm m\,s}^{-1}$. We did not detect any additional periodic signals or longer-term RV trends in this dataset for evidence of another RV planet.

\subsection{Floating Chunk Offset Method}\label{ssec:fco}
To check if there was any over-fitting GP analysis, we additionally performed a fit to the RVs using the floating chunk offset (FCO) method \citep[e.g.,]{Hatzes2010,Hatzes2011,Hatzes2014}. FCO is a common method for determining the RV semi-amplitudes of USPs (P $\lesssim$ 1 day), whose nightly RV variations are typically larger than individual RV measurement uncertainties \citep{Deeg2023}. By assigning nightly additive RV offsets to grouped observations, we focus entirely on intra-night RV variations and effectively remove any long-term (P $>$ 1 day) stellar and instrumental systematics. 

As done for the GP fit, RVs from single-measurement nights and nightly 3$\sigma$ outliers were excluded, noted with a (*) in Table \ref{tab:rv}. We assigned an offset term for each of 23 nights of observation, $\gamma_{night}$. This model differs from the GP model in that we remove the squared-exponential kernel and use only the white noise component in the covariance function. We implicitly include nightly offsets in the measured RV values in the likelihood function. We again used {\tt emcee} for sampling in an MCMC to explore the posteriors of the model, mirroring the procedure in our GP analysis. From the FCO method we detected TOI-6324 b at 2$\sigma$ precision, with a RV semi-amplitude of $K=2.19 \pm 1.02\;{\rm m\,s}^{-1}$. This is consistent with the result of the GP model within 1$\sigma$ uncertainties. Panel (d) of Fig.\ \ref{fig:rv} shows the Keplerian solution from the FCO model in phase with the GP model. Because the FCO model requires many more parameters (due to each $\gamma_{night}$) than the GP model, therefore introducing larger uncertainty, we chose to adopt the GP model result for further analyses. 

\section{Discussion} \label{sec:discussion}

\begin{figure}
    \centering
    \includegraphics[width=9cm]{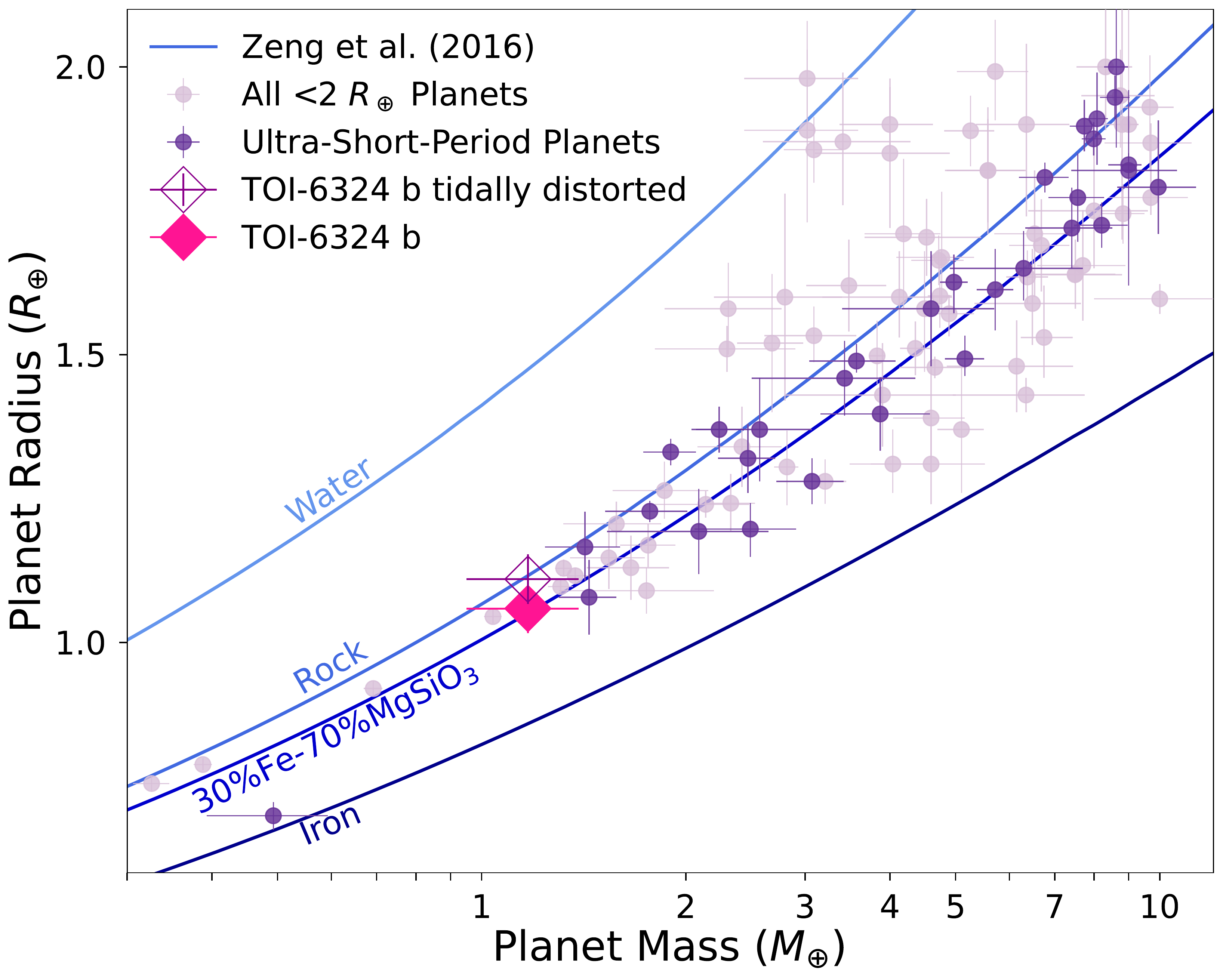}
    \caption{Mass-radius diagram of confirmed USPs (dark purple) and other rocky planets $<$ 2 R\e (light purple). TOI-6324 b is plotted in pink. The magenta-outlined point represents TOI-6324 b in the case that it is tidally disorted by 5\%. In the spherical case, CMF$_{R_{tran}}$=27$\pm37$\% while in the distorted case, CMF$_{R_{vol}}$=3$\pm40$\%.}
    \label{fig:mrd}
\end{figure}
\subsection{Planet Composition}\label{ssec:comp}
With precise mass and radius measurements (both $<$ 20\% uncertainties) of TOI-6324 b, we are able to approximate its interior composition. As mentioned, because USPs are so strongly irradiated by their host stars, it is suggested that any thick H/He envelopes should be stripped away \citep[e.g.,][]{sanchis-ojeda_study_2014,lundkvist_hot_2016,Wang2018,kreidberg_absence_2019,zhang_gj_2024}, allowing for direct constraint of their compositions. To compare with other confirmed USPs (queried from the NASA Exoplanet Archive\footnote{\url{https://exoplanetarchive.ipac.caltech.edu}}) and Earth-sized planets, we plotted TOI-6324 b on a mass-radius diagram along with the model composition curves from \citet{zeng2016}, shown in Fig.\ \ref{fig:mrd}. The light purple colored points represent all confirmed R$_{\rm p} < 2$ R\e\ planets with well-constrained masses and radii ($<$ 20\% uncertainties), while the dark purple points highlight the USPs among these with P $<$ 1 day. Similarly to other USPs, TOI-6324 b closely follows the Earth-like composition curve of 30\% iron and 70\% MgSiO$_3$ silicate rock.  

\subsection{Core-Mass Fraction}\label{ssec:core}
To further characterize the interior composition of TOI-6324 b we modeled the iron mass fraction (Fe-MF), which describes the fraction of the planet's mass that is comprised of the iron core. We modeled the core mass fraction (CMF) using {\tt Manipulate Planet}\footnote{\url{https://lweb.cfa.harvard.edu/~lzeng/manipulateplanet.cdf}}, a numerical tri-layer planetary interior model with an iron core, silicate rock mantle, and surface water layer \citep{zeng2013,zeng2016}. The model assumes an undifferentiated core and numerically solves the interior structure of a planet based on the Equation of State (EOS) extrapolated from Earth's Seismic Density Profile (using the Preliminary Reference Earth Model \citep[PREM;][]{PREM}). We adjusted the model to remove the water layer, and from this bi-layer iron and rock model we found a Fe-MF (adopted as the CMF in this model) of 27$\pm37$\% for TOI-6324 b. 

We additionally employed a more sophisticated model, {\tt exopie} \citep{Plotnykov2024}, based on {\sc superearth} \citep{valencia2006,Plotnykov2020}, which computes an iron mass fraction Fe-MF, distinct from the general CMF. The Fe-MF calculation takes into account the iron that could be in the mantle and removes any light elements in the core, as well as different degrees of core differentiation. We assumed that there is a variable amount of Fe in the mantle and Si in the core (0-20\%), but set the Mg/Si ratio in the mantle to be Earth-like ($\sim$0.9). Ni in the core was fixed to 10\%. Using a Monte Carlo scheme, we found that TOI-6324 b has a CMF=32$\pm19$\% and Fe-MF=30$\pm14$\%. These results are in agreement with the simpler model of \citet{zeng2016}. 

\subsection{Tidal Distortion} \label{ssec:dist}
The extremely short (6.7 hr) orbit of TOI-6324 b gives rise to the possibility that it is substantially tidally distorted. Such an effect has been suggested for TOI-6255 b, a 1.079 R\e\ USP on a 5.7 hr orbit \citep{Dai2024}. Tidal distortion would have a non-negligible effect on the volumetric radius $R_{\rm vol}$ and the effective transit radius of a planet, and thus must be accounted for when approximating the bulk planet composition. 

Following the method of \citet{Dai2024} we assess the degree of deformation due to tidal stress first by comparing the orbital period to the Roche limit of an incompressible fluid \citep{rappaport_roche_2013}. The Earth is not very well modeled as an incompressible fluid, so this formula is a very rough approximation:
\begin{equation}
    P_{\rm Roche}\approx 12.6 \,{\rm h}\, \bigg(\frac{\rho_{\rm p}}{1\,{\rm g\,cm}^{-3}}\bigg)^{-1/2}\,,
\end{equation}

\noindent where $\rho_{\rm p}$ is the mean bulk density of the planet. The orbital period of the Roche limit, $P_{\rm Roche}$, is the minimum orbital period that a planet may attain before being tidally disrupted. At periods shorter than $P_{\rm Roche}$, tidal forces imposed by the host star become stronger than the self-gravity of the planet, and the planet begins to disintegrate. For TOI-6324 b, $P_{\rm Roche}$ = 5.41 \p\ 0.15 hr, bringing it close to the tidal disruption limit with $P_{\rm orb}/P_{\rm Roche}$ = 1.24 \p\ 0.15, though not extremely so. Still, only TOI-6255 b rivals TOI-6324 b in approaching the tidal disruption limit among confirmed USPs (see Fig.\ 5 of \citealt{Dai2024}). The tidal decay timescale for TOI-6324 b, based on the star-to-planet mass ratio, mean stellar density, orbital period, and a nominal stellar tidal quality factor $Q^\prime_\star=10^7$, is $\tau_P\approx$ 8 Gyr. To reach the Roche Limit, the orbital period would need to decrease by $\sim$20\%, which would occur within $\tau_{Roche}\approx$550 Myr.

To estimate the tidal distortion $\delta R_{\rm p}$ of TOI-6324 b, we used the Love number $h_2$ for a solid homogeneous planet \citep{Love1944}, which describes the radial tidal displacement of a planet's surface. $h_2$ depends on the mean tensile strength of the planetary material. For Earth, $h_2$ has been estimated to be 0.6 to 0.9 \citep{lambeck1980}. We adopted $h_2=1$ assuming a weaker material strength for USPs, as in \citet{Dai2024}. This gives $\delta R_{\rm p}=0.07$ for TOI-6324 b. A more robust detection of ellipsoidal light variation than we could recover in the TESS phase curve (Sec.\ \ref{ssec:pc}, Fig.\ \ref{fig:pc}a) would provide more conclusive evidence that TOI-6324 b is tidally deformed. 

Since TOI-6324 b, like other USPs, is expected to be tidally locked \citep{winn_kepler-78_2018}, its planetary rotation period is the same as its orbital period of 6.7 hr. This may also lead to a non-negligible rotational deformation. This is quantified by
\begin{equation}
    q = \frac{\Omega^2R_p^3}{GM_p}\,,
\end{equation}
\noindent or the surface gravity acting against centrifugal forces at the planet's surface. We estimated $q=0.04$ for TOI-6324 b. The deformation of the planet due to the combination of tidal bulge and rotational deformation results in an ellipsoidal shape with semi-major axes $R_1$ pointing towards the host star, $R_2$ pointing along the direction of orbital motion, and $R_3$ pointing along the planet's rotation axis \citep{correia2014}. The total volumetric radius of the planet is defined as $R_{vol}\equiv(R_1R_2R_3)^{1/3}$. During transit, a planet's transit radius is roughly $R_{tran}\equiv(R_2R_3)^{1/2}$, and the true volumetric radius is larger than the transit radius by $R_{vol}/R_{tran}=1+\frac{7}{12}\delta R_{\rm p}$ \citep{Dai2024}. We found $R_{vol}/R_{tran} = 1.05$, i.e., there is a 5\% increase in the volumetric radius compared to the transit radius. The modeled CMF changes significantly from CMF$_{R_{tran}}$=27$^{+37}_{-27}$\% to CMF$_{R_{vol}}$=3$\pm40$\%, more consistent with a pure-rock composition. However, within the large uncertainties, this qualitatively does not alter the results of Sections \ref{ssec:comp} and \ref{ssec:core} (Fig.\ \ref{fig:mrd}). 

\subsection{JWST Prospects}
USPs such as TOI-6324 b are particularly excellent targets for phase curve and secondary eclipse observations with JWST. Several such systems have been selected or have already been successfully observed with JWST (e.g., 55 Cnc e, GJ 367 b, K2-141 b, LHS 3844 b, TOI-561 b, TOI-1685 b, TOI-2445 b, TOI-4481 b, and WASP-47 e). Precise study of their surfaces will revolutionize our understanding of planetary atmospheres and geology. USPs tend to orbit late-type stars \citep{Dai2024}, and thus their flux contrasts with their host stars are observationally favorable, on the order of 100 ppm in the infrared. We computed the Emission Spectroscopy Metric \citep[ESM;][]{kempton2018} for TOI-6324 b to be 25. This is the highest ESM target among Earth-sized (0.8 R\e\ $<$ R$_p$ $<$ 1.2 R\e) planets. The TESS data alone are insufficient for robustly constraining the phase curve variation, though we do find preliminary evidence of several tens of ppm in the optical ($ A_{ill}=42\pm28$ ppm and $A_{ELV}=21\pm17$ ppm; Sec.\ \ref{ssec:pc}, Fig.\ \ref{fig:pc}a). For USPs, the planet-to-star flux ratio is expected to be much more significant in the mid-IR \citep{Dai2024}. In the MIRI/LRS bandpass (5-12 $\mu$m), the planet's thermal emission increases to $\sim$280 ppm assuming a low Bond albedo.  We aim to include more detailed modeling of the TESS phase curve variation in a future paper with comparison to JWST observations. 

There are several scientific avenues for detailed characterization of TOI-6324 b with JWST MIRI/LRS \citep{kendrew2015} observations. Since USPs are tidally locked, they will have constant daysides and nightsides, resulting in a significant longitundal temperature contrasts across their surfaces. An observed offset in the phase curve would signify heat re-circulation over the surface of the planet due to the presence of an outgassed secondary atmosphere \citep[e.g.,][]{Knutson2009,demory2013,showman2015,vp2016,angelo2017}. A lack of phase offset indicates a bare rock surface \citep[e.g.,][]{zhang_gj_2024,mansfield2024}. In this case, variations in the phase curve will trace the projected shape of the planet across its orbit and help to constrain the extent of tidal distortion and provide key insights into the tidal interactions USPs undergo with their host stars (see Sec.\ \ref{ssec:dist}; \citealt{Dai2024}). 

For a bare rock surface, the emission spectrum will constrain the dominant surface rock type of the planet \citep{hu2012,kreidberg_absence_2019,whittaker2022,zhang_gj_2024}. The approximate equilibrium temperature of TOI-6324 b is $T_{eq}=1216\pm60$ K (Table \ref{tab:params}), assuming a low albedo of 0.1. This is below the zero-pressure melting point of common silicates found on Earth (such as peridotite at 1390 K), and thus it is unlikely that TOI-6324 b has a pure lava surface, although the hotter dayside, which receives constant irradiation may be partially molten. Assuming zero albedo, the temperature at the substellar point would be approximately 1824 K, which is sufficient to melt silicates. Excessive tidal heating may also contribute to violent volcanism at the surface \citep{jackson2008,hu2012}, if the planet actually has a non-zero eccentricity, contrary to our assumption in Sec \ref{ssec:transit}. Detection of basaltic or ultramafic surface mineralogy, based on the Si-O absorption feature from 8-10 $\mu$m, will constrain the ongoing geologic processes. Ultramafic rock is formed from high-temperature volcanism and has strong Si-O features, while basaltic rock is formed from Earth-like extrusive volcanism and typically exhibits weaker Si-O features. The presence of granitoid rock (the primary component of Earth's continental crust) is unlikely at the extreme temperatures of TOI-6324 b, which are high enough to melt feldspar and quartz (the primary minerals in granite). Dedicated simulations of these observations and effects will be included in future works. 

\section{Summary} \label{sec:sum}
In summary, we report the confirmation of TOI-6324 b using TESS transit observations and RV monitoring with KPF. The key findings are as follows: 
\begin{itemize}
    \item TOI-6324 b has a transit radius of 1.059 \p\ 0.041 R\e. This is one of the closest to Earth-sized USP confirmed to date, and one of the only Earth-sized USPs with a precise mass measurement. 
    \item TOI-6324 b has a mass of 1.17 \p\ 0.22 M\e. The RV mass constraints from both Gaussian Process Regression and Floating Chunk Offset methods are consistent within 1$\sigma$. 
    \item The 27$\pm37$\% CMF of TOI-6324 b, as with other $\gtrsim$ 1 R\e\ USPs, is consistent with an Earth-like composition. 
    \item Accounting for tidal distortion due to its short orbit ($P_{\rm orb}/P_{\rm Roche}$ = 1.24 \p\ 0.15), the radius may be inflated by $\sim5$\%. This would alter the CMF to 3$\pm40$\%. 
    \item We report preliminary evidence of phase curve variation ($A_{ill}=42\pm28$ ppm) from the TESS data. With an ESM of 25, TOI-6324 b is among the most favorable USP targets for JWST phase curve and emission spectrum observations to constrain its surface mineralogy and tidal deformation. 
\end{itemize}

\begin{center}
    ACKNOWLEDGEMENTS
\end{center}

\textit{Software:} {\tt arviz} \citep{exoplanet:arviz}, {\tt astropy} \citep{exoplanet:astropy13, exoplanet:astropy18}, {\tt batman} \citep{Kreidberg2015}, {\tt dynesty} \citep{Speagle2020,sergey_koposov_2024_12537467}, {\tt emcee} \citep{emcee2013}, {\tt exopie} \citep{Plotnykov2024}, {\tt exoplanet} \citep{exoplanet:joss,exoplanet:zenodo}, {\tt isoclassify} \citep{huber2017}, {\tt Lightkurve} \citep{lightkurve}, {\tt lmfit} \citep{lmfit}, {\tt Manipulate Planet} \citep{zeng2013,zeng2016}, MESA Isochrones and Stellar Tracks \citep[MIST;][]{MIST}, {\tt pyMC3} \citep{exoplanet:pymc3}, {\tt radvel} \citep{fulton_radvel_2018}, {\tt SpecMatch-Emp} \citep{Yee2017}, {\tt starry} \citep{exoplanet:luger18}, {\sc superearth} \citep{valencia2006}, {\tt theano} \citep{exoplanet:theano}, {\tt wōtan} \citep{hippke_wotan_2019}

\textit{Facilities:} Keck I/KPF, Keck I/HIRES, Keck II/NIRC2, \textit{TESS}

\textit{Data:} All the \textit{TESS} data used in this paper can be found in MAST: \dataset[10.17909/t9-nmc8-f686]{http://dx.doi.org/10.17909/t9-nmc8-f686}. This work uses data supplied from the NASA Exoplanet Archive: \dataset[10.26134/ExoFOP3]{http://dx.doi.org/10.26134/ExoFOP3}

The authors wish to recognize and acknowledge the very significant cultural role and reverence
that the summit of Mauna Kea has always had within the indigenous Hawaiian community. We
are most fortunate to have the opportunity to conduct observations from this mountain.

RAL acknowledges this material is based upon work supported by the National Science Foundation Graduate Research Fellowship Program under Grant No. 1842402 and Grant No. 2236415. Any opinions, findings, and conclusions or recommendations expressed in this material are those of the author(s) and do not necessarily reflect the views of the National Science Foundation.

We thank Ellen Price for helpful discussions about tidal distortion. 

A NASA Key Strategic Mission Support titled ``Pinning Down Masses of JWST Ultra-short-period Planets with KPF'' (PI: F. Dai) provided the telescope access and funding for the completion of this project. 

This work was supported by a NASA Keck PI Data Award, administered by the NASA Exoplanet
Science Institute. Data presented herein were obtained at the W. M. Keck Observatory from
telescope time allocated to the National Aeronautics and Space Administration through the
agency's scientific partnership with the California Institute of Technology and the University of
California. The Observatory was made possible by the generous financial support of the W. M.
Keck Foundation.

DRC acknowledges partial support from NASA Grant 18-2XRP18\_2-0007. 

JMJO acknowledges support from NASA through the NASA Hubble Fellowship
grant HST-HF2-51517.001, awarded by STScI. STScI is operated by the
Association of Universities for Research in Astronomy, Incorporated,
under NASA contract NAS5-26555.

LMW acknowledges support from the NASA Exoplanet Research Program (grant no. 80NSSC23K0269). 

N.S. acknowledges support by the National Science Foundation Graduate Research Fellowship Program under Grant Numbers 1842402 \& 2236415 and the National Aeronautics and Space Administration (80NSSC21K0652).

This research was carried out in part at the Jet Propulsion Laboratory, California Institute of Technology, under a contract with the National Aeronautics and Space Administration (80NM0018D0004).

This research was carried out, in part, at the Jet Propulsion Laboratory and the California Institute of Technology under a contract with the National Aeronautics and Space Administration and funded through the President’s and Director’s Research \& Development Fund Program.

This research has made use of the Exoplanet Follow-up Observation Program (ExoFOP; DOI: 10.26134/ExoFOP5) website, which is operated by the California Institute of Technology, under contract with the National Aeronautics and Space Administration under the Exoplanet Exploration Program. 

We acknowledge the use of public TESS data from
pipelines at the TESS Science Office and at the TESS
Science Processing Operations Center. Resources supporting this work were provided by the NASA HighEnd Computing (HEC) Program through the NASA
Advanced Supercomputing (NAS) Division at Ames Research Center for the production of the SPOC data products.

This paper made use of data collected by the TESS
mission and are publicly available from the Mikulski
Archive for Space Telescopes (MAST) operated by the
Space Telescope Science Institute (STScI).
Funding for the TESS mission is provided by NASA’s
Science Mission Directorate. KAC and CNW acknowledge support from the TESS mission via subaward s3449
from MIT.

Some of the data presented herein were obtained at Keck Observatory, which is a private 501(c)3 non-profit organization operated as a scientific partnership among the California Institute of Technology, the University of California, and the National Aeronautics and Space Administration. The Observatory was made possible by the generous financial support of the W. M. Keck Foundation.


\bibliography{bib_extra}{}
\bibliographystyle{aasjournal}



\end{document}